%% file: avgmaxfidelity_main.tex
\documentclass[american,letterpaper,nofootinbib,aps,prl,reprint,tightenlines,superscriptaddress,twocolumn,twoside,longbibliography,dvipsnames]{revtex4-1}
\pdfoutput=1

\usepackage{XCharter}
\usepackage[T1]{fontenc}
\usepackage{mathptmx}
\usepackage[framemethod=tikz]{mdframed}
\usepackage{amsmath,amsthm,amssymb,mathrsfs}
\usepackage{slashed}
\usepackage{graphicx}
\usepackage[export]{adjustbox}
\usepackage{float}
\usepackage{multirow, makecell}
\usepackage{hyperref}
\usepackage{enumitem}
\hypersetup{colorlinks=true,citecolor=red,linkcolor=NavyBlue,urlcolor=NavyBlue}
\usepackage[caption=false]{subfig}
 
\usepackage{natbib}

\usepackage{relsize}
\usepackage{footmisc}
\usepackage[left=1.55cm,right=1.55cm,top=1.6cm,bottom=1.56cm]{geometry}
\usepackage{tcolorbox}
\usepackage{longtable}

\begin{document}

\title{Teleportation fidelity of quantum repeater networks}

\author{Ganesh Mylavarapu}
\thanks{Equal contribution}
\email{ganesh.mylavarapu@research.iiit.ac.in}
\affiliation{Center for Security, Theory and Algorithmic Research, 
International Institute of Information Technology, Hyderabad 500 032, India}

\affiliation{Centre for Quantum Science and Technology,
International Institute of Information Technology, Hyderabad 500 032, India.}

\author{Subrata Ghosh}
\thanks{Equal contribution}
\email{jaba.subrata94@gmail.com}
\affiliation{Center for Computational Natural Science and Bioinformatics, 
International Institute of Information Technology, Hyderabad 500 032, India}
\affiliation{Division of Dynamics,  Lodz University of Technology, Stefanowskiego 1/15, 90-924  Lodz, Poland}

\author{Chittaranjan Hens}
\email{chittaranjan.hens@iiit.ac.in}
\affiliation{Center for Computational Natural Science and Bioinformatics, 
International Institute of Information Technology, Hyderabad 500 032, India}

\author{Indranil Chakrabarty}
\email{indranil.chakrabarty@iiit.ac.in}
\affiliation{Centre for Quantum Science and Technology,
International Institute of Information Technology, Hyderabad 500 032, India.}
\affiliation{Center for Security, Theory and Algorithmic Research, 
International Institute of Information Technology, Hyderabad 500 032, India}

\author{Subhadip Mitra}
\email{subhadip.mitra@iiit.ac.in}
\affiliation{Center for Computational Natural Science and Bioinformatics, 
International Institute of Information Technology, Hyderabad 500 032, India}

\affiliation{Centre for Quantum Science and Technology,
International Institute of Information Technology, Hyderabad 500 032, India.}

\begin{abstract} 
\noindent 
We show that the average of the maximum teleportation fidelities between all pairs of nodes in a large quantum repeater network is a measure of the resourcefulness of the network as a whole. {It can be used to rank networks of arbitrary topologies, with and without loops. For demonstration, w}e use simple Werner state-based models to characterise some fundamental {mainly} loopless) topologies (star, chain, some trees, {and also ring and the complete graph}) with this measure in three (semi)realistic scenarios. Most of our results are analytic and apply to arbitrary network sizes. We identify the parameter ranges where these networks can achieve quantum advantages and show the large-$N$ behaviours.
\end{abstract}
\maketitle

\noindent \textit{Introduction:} In the future, quantum entanglement-based networks are expected to perform various computing and information-processing tasks in distributed scenarios~\cite{aaronson2011computational,menicucci2006universal,knill2001scheme,cirac1999distributed,fitzi2001quantum,giovannetti2008quantum,das2021practically, bollinger1996optimal,giovannetti2006quantum,bennett1993teleporting,horodecki1996teleportation,bennett1992communication,
pati2000minimum,bennett2005remote,pati2023teleportation,ekert1991quantum,shor2000simple,hillery1999quantum,singh2024controlled,ray2016sequential,pironio2010random,yang2019distributed,sazim2015retrieving,sazim2013study,sadhu2023practical}, ultimately leading to the quantum internet~\cite{dowling2003quantum,kimble2008quantum,Google,Rigeti}. While large classical networks are known to show many intriguing features~\cite{albert2002statistical,watts1998collective,pastor2015epidemic,gao2016universal,boccaletti2014structure,ji2023signal,hens2019spatiotemporal,ghavasieh2024diversity,meena2023emergent,moore2020predicting}, large-scale quantum networks remain largely unexplored. Theoretically, they raise several questions, including fundamental questions about non-locality~\cite{tavakoli2022bell,branciard2012bilocal,branciard2010characterizing,mukherjee2023persistency}. Various issues, like the role of network topology on quantum key distribution~\cite{chen2009field}, entanglement percolation~\cite{cuquet2009entanglement}, etc., are getting attention in the recent literature. On the practical side, one has to understand the pros and cons of different types of quantum networks before deploying over large areas. For example, the network can use satellite-based technology~\cite{aspelmeyer2003long,bedingtonprogress} or be ground-based~\cite{briegel1998quantum,duan2001long}. For distant ground-based communication, one normally has to transfer an entangled qubit physically~\cite{sazim2013study,mylavarapu2023entanglement,gour2004remote,gour2005family,azuma2023quantum,sangouard2011quantum}, which is prone to loss of entanglement (unless one uses robust distillation protocols). However, intermediate repeater stations can establish entanglement between a widely separated source and target pair via entanglement swappings and transfer quantum information~\cite{briegel1998quantum,duan2001long,zukowski1993event,bose1998multiparticle,meignant2019distributing,winnel2022achieving}. In this letter, we focus on quantum networks established through repeater stations performing Bell-state measurements with free local operations and classical communication.

We know that, a priori, not all entangled states are useful as resources for quantum protocols~\cite{horodecki1996teleportation,chakrabarty2010teleportation,adhikari2008quantum,chakrabarty2011deletion} or show quantum advantages. For example, in the case of quantum teleportation~\cite{bennett1993teleporting}---the protocol to transfer quantum information---the {\it maximum} achievable fidelity (obtained by performing the Bell measurement resulting in the maximum fidelity) of a two-qubit resource state $\rho$ is $F_\rho^{\rm max}=(1+\mathcal{N}(\rho)/3)/2$ with $\mathcal{N}(\rho)=Tr(\sqrt{T^{\dagger}T})$, where $T$ is the correlation matrix of $\rho$~\cite{horodecki1996teleportation}. This implies that unless $\rho$ is maximally entangled (ME), $F_\rho^{\rm max}<1$. The state $\rho$ shows quantum advantage only if $F_\rho^{\rm max}>2/3$, the maximum that can be achieved without using entangled states. 
{In general, a teleportation witness~\cite{ganguly2011entanglement} can identify states useful for teleportation. (Interestingly, after local filtering, all entangled states become useful for teleportation~\cite{li2021activating}.)}
However, when entangled qubits are shared among many parties to form large teleportation networks, numerous pathways for information transfer open up. Although, theoretically, we can assume all links (shared states) in a large network are maximally entangled (i.e., they have $F_\rho^{\rm max}=1$, in which case, the maximum teleportation fidelity of the entire network is trivially one), the presence of factors like noise will make such ME networks highly challenging to realise in practice. Hence, in a realistic scenario, we will need a measure to quantify the achievable fidelity of a network as a whole. 

Here, we use Werner states to model large quantum repeater networks with fundamental topologies (stars, chains, and some trees with the same number of links) and show that the average of the maximum teleportation fidelities ($F^{\rm max}_{\rm avg}$)---the highest teleportation fidelity one can achieve between a source and a target averaged over all source and target combinations in a network---can be used as a measure to compare networks' teleportation abilities (i.e., it can act as a quantifier of the resourcefulness of a teleportation network as a whole). We consider some realistic scenarios and obtain $F^{\rm max}_{\rm avg}$  analytically for arbitrary network size in each case. Our results show that $F^{\rm max}_{\rm avg}$ can rank networks; it is maximum for the star and minimum for the chain for identical parameters. We also identify the parameter ranges for which a large network shows quantum advantages. Our results characterise quantum networks with respect to a specific task (teleportation) and establish the threshold values for quantum advantages (resourcefulness) in 
quantum networks. {The importance of identifying global measures of quantum networks can hardly be overemphasised, given the recent theoretical and experimental advancements in teleporting qubits over quantum networks~\cite{hermans2022qubit,ghosal2024repeater}. (See, e.g., Ref.~\cite{centrone2023cost} for a recent topology-based study on global measures of quantum networks in continuous-variable systems.)} It is important to mention here that while there are universal limitations on how much quantum communication is possible over networks, memory effects can be used to bypass those~\cite{das2021universal}.  

{Before we move on to discuss our models, some observations are in order. Repeater-based technologies need not be restricted to two-qubit states only but can be extended to multi-qubit systems (like $W$ states~\cite{miguel2023quantum}). Similarly, they can also be used in continuous-variable cluster state-based networks \cite{asavanant2019generation} and networks based on orbital angular momentum optical modes \cite{wang2020large}.}
\medskip

\begin{figure*}
    \centering
    {\includegraphics [width=0.8\columnwidth]{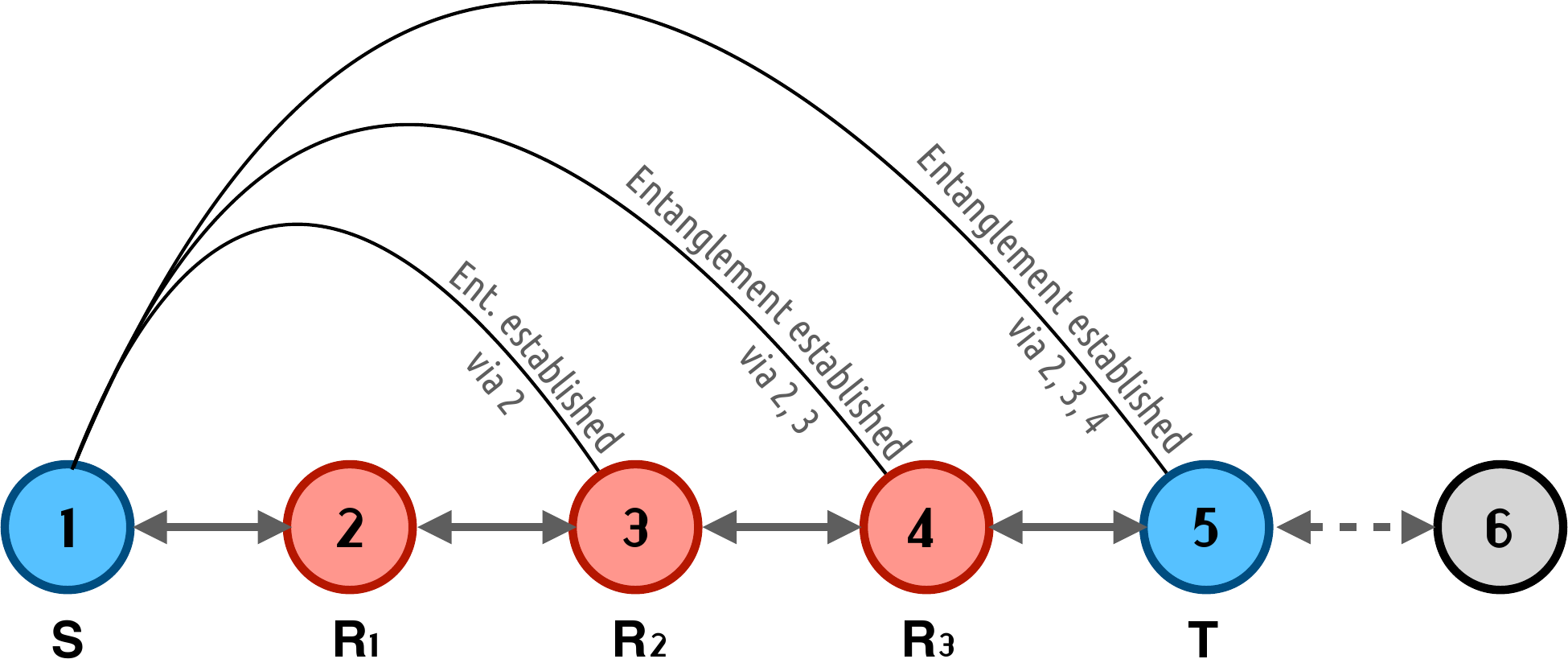}\hspace{1cm}
    \includegraphics [width=0.4\columnwidth]{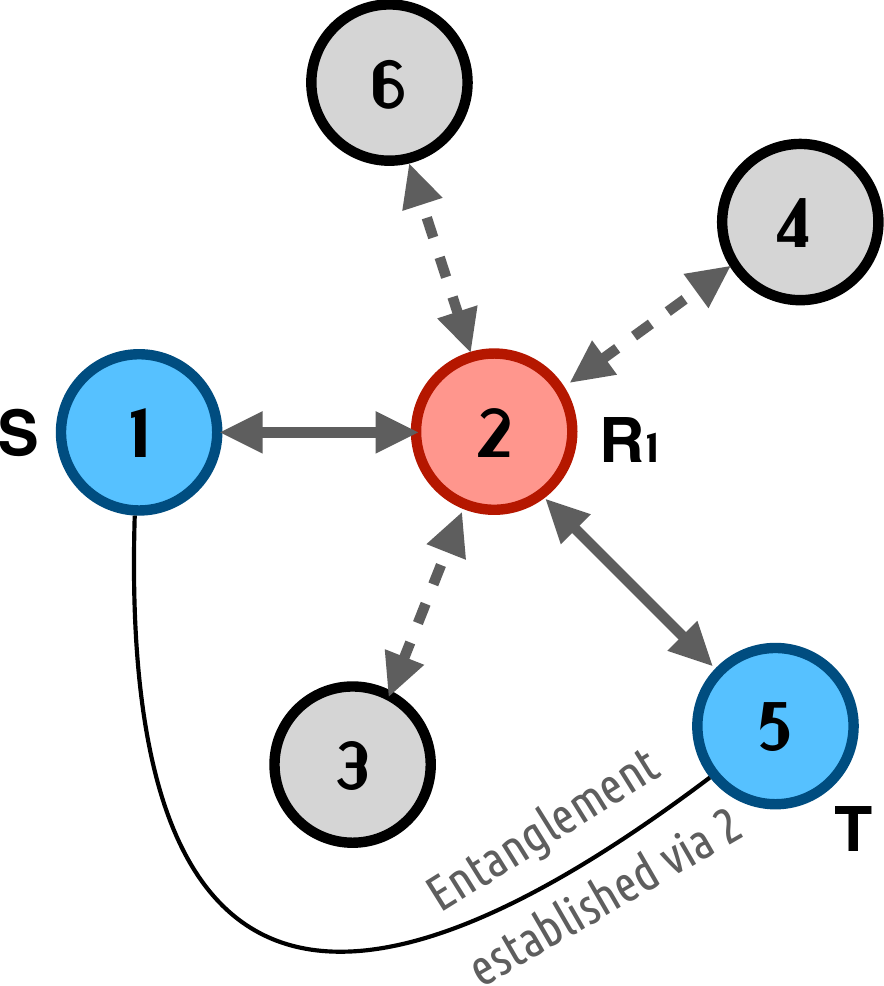}\hspace{1cm}
    \includegraphics [width=0.4\columnwidth]{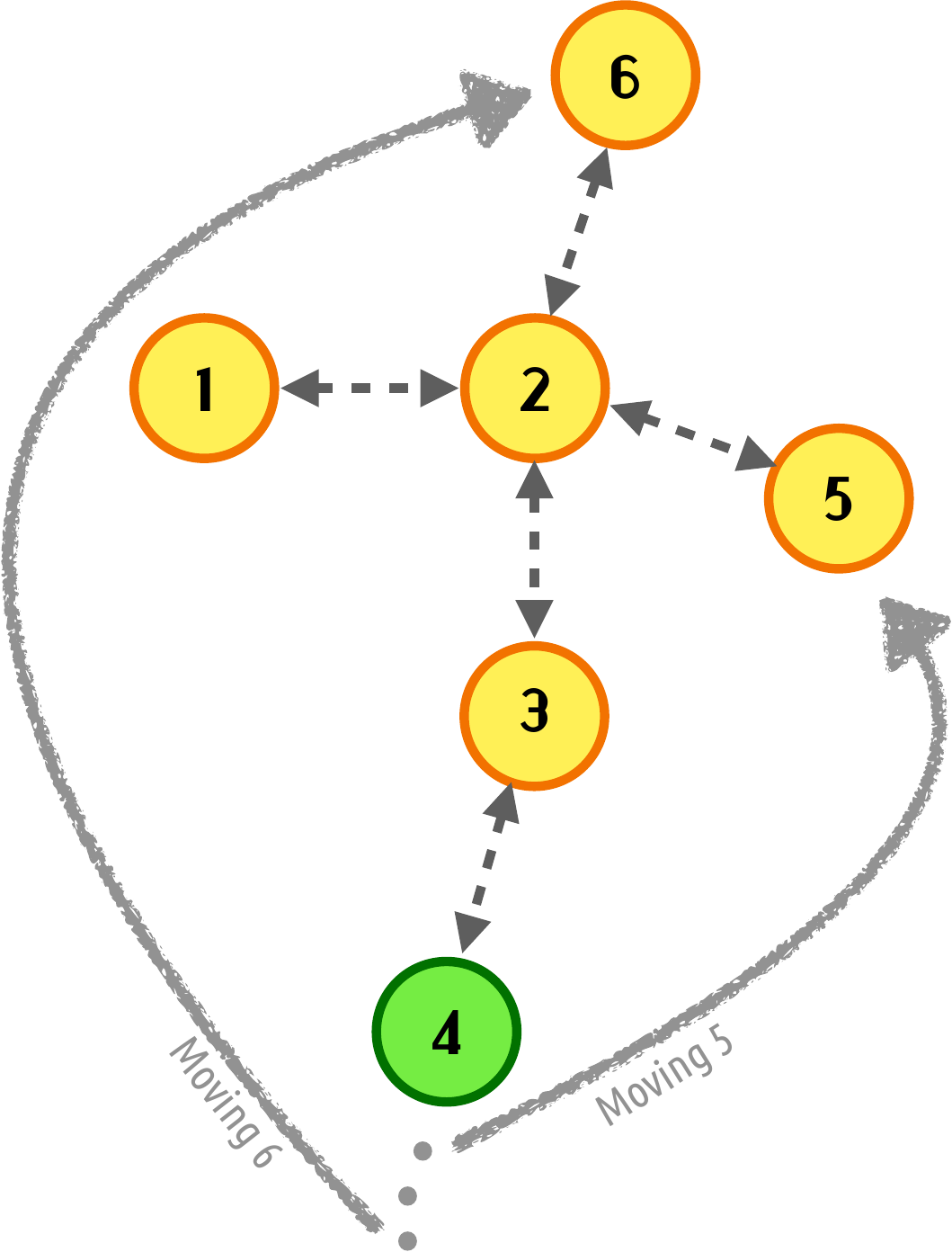}
    }
    \caption{\label{fig:entchain} Quantum repeater networks with $N=6$ nodes (stations) and $L=5$ links (shared states): the chain (left), the star (middle), and the second intermediate flower (right). The nodes are connected by ensembles of Werner states (see text). To establish entanglement between Node $1$ (source) and Node $5$ (target) (shown in blue) the intermediate repeater stations ($R_i$, in red) perform entanglement swappings. Intermediate flowers are some specific tree networks. They can be obtained by taking nodes from a side of the chain and joining them with the second node on the opposite side (see text). The petals are shown in yellow and the stem in green.}
    \label{fig:fig1}
\end{figure*}

\noindent \textit{Models:} 
To model 
quantum repeater networks in a simple and calculable manner, we consider $N$-node networks with $L$ undirected links made up of Werner states ~\cite{WernerPhysRevA.40.4277} parameterised by weight factors, $p_i$.\footnote{In practice, an intermediate link will be destroyed after a swap. Hence, such networks require an ensemble of Werner states between two consecutive nodes.} {Since topologies with loops are more complex and subtle, as the first step, we mainly focus on examples from loopless topologies for clarity and simplicity in this letter. However, our definition of $F^{\rm max}_{\rm avg}$ is general and applicable for loops as well. We touch upon the topic of loops at the end and discuss the cases of some simple but fundamental loop topologies in Appendix~\hyperref[sec:loops]{B}.} 

For fixed numbers of nodes and links, the distribution of a network's neighbouring nodes (degree) varies from graph to graph. For instance, a star has the highest maximum degree---it has a hub with all $L$ links directly connected to it, whereas a simple chain has the lowest maximum degree: $2$ (see  Fig.~\ref{fig:entchain}). Between these two extremes, there are intermediate trees with a maximum degree between $3$ and $N-2$ obtained by rearranging the nodes. For our purpose, we only focus on some specific trees. To get these specific topologies, we can start (for example) from a chain and cut the link at one side (say, the link between the last node on the right and the one before) and link the loose node with the second node on the other side. If we keep repeating this step, we get these intermediate shapes and, finally, the star in $L-2$ steps. In other words, all these intermediate trees have the structure of a chain connected to one of the outer nodes of a smaller star, i.e., like the petals in a flower connected to a stem [see Fig.~\ref{fig:entchain} (right)]. We refer to these specific structures as intermediate flowers (IFs). (It is, of course, possible to construct other types of trees by rearranging them differently. However, it is enough to consider these special ones for the present purpose since their $F^{\rm max}_{\rm avg}$ values will be bounded by those of the star and the chain.) 

In the case of quantum repeater networks made of Werner states, the maximum teleportation fidelity through a particular path ($\mathcal P$) connecting the source ($S$) and the target ($T$) can be calculated analytically as~\cite{mylavarapu2023entanglement,sen2005entanglement} $F^{\rm max}_{{\rm ST},\mathcal P}(\rho_{\rm wer}) = \left(1 + \prod_{i\in\mathcal P} p_{i}\right)/2$. 
If the intermediate links in $\mathcal P$ are all ME (i.e., $p_i=1~\forall i\in \mathcal P$), $F^{\rm max}_{{\rm ST},\mathcal P}(\rho_{\rm wer}) =1$. On the other hand, the path will not show any quantum advantage (i.e., behave no better than a classical connection) if $p_i\to0$. If S and T are connected via multiple paths {(e.g., in networks with loops)}, let $\mathcal P_{\rm max}$ be the path(s) with the maximum fidelity. We get the average highest-achievable teleportation fidelity if we take the average of $F^{\rm max}_{{\rm ST},\mathcal P_{\rm max}}(\rho_{\rm wer})$ over all possible combinations of S and T (i.e., any pair of nodes can be the source and the target):
\begin{align}
    F^{\rm max}_{\rm avg}(\rho_{\rm wer}) =&\ \langle F^{\rm max}_{{\rm ST},\mathcal P_{\rm max}}(\rho_{\rm wer})\rangle_{{\rm ST}}\nonumber\\
    =&\ \langle F^{\rm max}_{\mathcal P}(\rho_{\rm wer})\rangle_{\mathcal P} \quad ({\rm loopless}), \label{eq:fmaxavgdef}
\end{align}
where the last expression follows from the fact that in the absence of loops, the path between any S and T pair is unique. Hence, in the loopless case, averaging over S and T pairs is equivalent to averaging over all possible paths in the network. {However, as mentioned,  $F^{\rm max}_{\rm avg}(\rho_{\rm wer})$ defined in Eq.~\eqref{eq:fmaxavgdef} is general and is applicable to networks with loops as well (see Appendix~\hyperref[sec:loops]{B} for illustrations).}

The above discussion shows that the simple Werner states-based models let us parameterise the network fidelities with a simple parameter set $\{p_i\}$. A large quantum network as a whole is expected to show quantum advantage if $F^{\rm max}_{\rm avg}>2/3$. This is because, without entangled states, each path can only achieve a maximum teleportation fidelity of $2/3$. Hence, at that threshold, the average maximum fidelity also becomes $2/3$. However, since this is only true on average, one can also look for the lowest value for which at least one path in the network shows quantum advantage ($F^{\rm max}_{\rm ST}>2/3$ for one or more paths). Similarly, one can consider the $F^{\rm max}_{\rm avg}$ value for which all paths in the network show quantum advantages.

To characterise the network parameters at these values, we estimate $F^{\rm max}_{\rm avg}$ in some representative scenarios: (A) all $p_i=p$ where $0\leq p<1$; (B) $p_i\in\{p,1\}$, i.e., a fraction of the links are ME and all the others have $p_i=p$; and (C) the $p_i$'s are randomly sampled from the uniform distribution, $\mathcal U_{[0,1]}$. We show analytic results for the first two cases---the first one is parametrised by $N$ and $p$, and the second one is parametrised by $N$, $p$, and $M$, the number of ME links (or $m$, the fraction of ME links).
\medskip

\noindent \textit{Scenario A:} For a quantum star network of $N$ nodes and $L=N-1$ links we have
\begin{align}
    \left.F^{\rm max}_{\rm avg}(N,p)\right|_{\rm star} 
    =&\ \left({^{L}\!C_1}\mathcal F_1 +  {^{L}\!C_2}\mathcal F_2\right)/(^{N}\!C_2),
    \label{eq:p_star}
\end{align}
where $\mathcal F_n \equiv (1+p^n)/2$. Since, in this case, all links have the same weight $p$, we can understand this relation by simply measuring the path length between any pair of nodes in the units of the number of links. Out of the $^{N}\!C_2$ possible paths, $N-1$ have length one (hence each contributes as $(1+p)/2$ to the sum of the highest-achievable fidelities, $F^{\rm max}_{\rm tot}$, as shown in the numerator) and the rest $^{(N-1)}\!C_2$ have length two (each contributes as $(1+p^2)/2$ to $F^{\rm max}_{\rm tot}$). 

For a chain with the same number of nodes and links as the star, we have
\vspace{-0.2cm}
\begin{equation}
 \left.F^{\rm max}_{ \rm avg}(N,p)\right|_{\rm chain}=\frac{1}{^{N}\!C_2}\left[\sum_{\ell=1}^{L}(N-\ell)\mathcal F_\ell\right].  
  \label{eq:p_chain}
\end{equation}
Again, it is easy to see that there are $L-\ell+1$ paths of length $\ell$ contributing to $F^{\rm max}_{\rm tot}$ as $\mathcal F_\ell$. We notice that since $p<1$, each term contributes less and less with increasing $\ell$, i.e., smaller paths contribute more. For very high $\ell$, $\mathcal F_\ell\approx 1/2$, and hence, every long path contributes to $F^{\rm max}_{ \rm avg}$ as $1/2$ in the numerator and $1$ in the denominator.

The $k^{\rm th}$ IF (obtained after transferring $k$ nodes from the chain to the star) can be thought of as a star of $(k+2)$ links [or $(k+3)$ nodes] plus a chain of $(L-k-2)$ links [or $(N-k-2)$ nodes] with one common node. In this case, we have
\vspace{-0.1cm}
\begin{align}
 &\left.\hspace{-0.1cm}F^{\rm max}_{ \rm avg}(N,p)\right|_{{\rm flower}_k}\nonumber\\
 &=\ \frac{1}{^{N}\!C_2}\Bigg[\Bigg\{{^{k+2}\!C_1}\mathcal F_1 +  {^{k+2}\!C_2}\mathcal F_2\Bigg\}
 + \left\{\sum_{\ell=1}^{L-k-2}(N-k-2-\ell)\mathcal F_\ell\right\}\nonumber\\
 &\qquad \qquad +\left\{
 \sum_{\ell=1}^{L-k-2}\left((k+1)\mathcal F_{(\ell+2)}+\mathcal F_{(\ell+1)} \right)\right\}
 \Bigg]\nonumber\\
 &=\ \frac{1}{^{N}\!C_2}\Bigg[{^{k+1}\!C_2}\mathcal F_2
 +\sum_{\ell=1}^{L-k}(N-\ell)\mathcal F_\ell\Bigg].
  \label{eq:kthif}
\end{align}
Here, the first and second sets of terms in the first line of the numerator come from the star and the chain, respectively, and the third set comes from overlapping paths connecting these two structures. For the one shown in the middle of Fig.~\ref{fig:entchain}, $k=2$ and $N=6$; hence it has $F^{\rm max}_{ \rm avg}(6,p)=(5\mathcal F_1+7\mathcal F_2+3\mathcal F_3)/15$, as expected.
\medskip

\noindent \textit{Scenario B:} In this scenario, any $M$ out of the $L$ links are ME; the rest have $p_i=p$. {(As a practical motivation for considering this scenario, we can imagine a situation where one starts with all $p$-Werner states and successfully distils~\cite{clarisse2006entanglement} $M$ of those to create maximally entangled links.)} Since the maximum teleportation fidelity of a ME link is one ($=\mathcal F_0$), the presence of a ME link does not affect the achievable fidelity of a path of length more than one, i.e., we can ignore the ME links while measuring the path length in terms of the number of $p$-links in it. 
We get
\begin{align}
    &\hspace{-0.8cm}\left.F^{\rm max}_{\rm avg}(N,M,p)\right|_{\rm star} \nonumber\\
    =&\ \frac{1}{^{N}\!C_2}[{^{M+1}\!C_2}\mathcal F_0 + (M+1)(L-M)\mathcal F_1 + {^{L-M}\!C_2}\mathcal F_2].
     \label{p_star_m}
\end{align}
Since the $M$ ME links can be placed in ${^L\!C_M}$ ways the total number of paths is not ${^N\!C_2}$ but ${^N\!C_2}{^L\!C_M}$ in this case. However, since the links in the star are all similarly connected to the hub, the ${^L\!C_M}$ factor cancels out in the average.  

For the chain with $M$ ME links, we get
\begin{align}
&\hspace{-1cm}\left.F^{\rm max}_{\rm avg}(N,M,p)\right|_{\rm chain} = \frac{N}{(N+1-M)}\nonumber\\
    &\ \times \frac{1}{^N\!C_2}\Bigg[\frac{(N+1)}{(N-M)} \sum\limits_{\ell = 1}^{L-M}(N-M-\ell)\mathcal F_\ell  +M\mathcal F_0\Bigg].
    \label{p_chain_m}
\end{align}
As earlier, we have factored out ${^L\!C_M}$ from the numerator. One can obtain this result intuitively by considering a problem of binary string arrangements: let us represent each ME link by a $0$ and each $p$-link by a $1$. We start with a bag of $M$ zeros and $L-M$ ones and count the possible arrangements of binary strings of length $1\leq \ell\leq L$. The number of $p$-links of a string can be calculated easily by adding the digits in a string ($=$ the total number of $1$'s). 

The expression for an arbitrary IF is lengthier but can be derived similarly. We show it in Appendix~\hyperref[sec:appscenarioB]{A}. 

Before presenting the numerical results, we look at an interesting relation. We can consider the average of the effective path lengths ($\sim$ resistance distances~\cite{Klein1993}) in a network, $\ell^p_{\rm avg}$, measured in terms of the number of non-ME links ($p$-links $\sim$ resistors), i.e., without counting the ME links ($\sim$ zero-resistance). If all $p_i=p<1$ (as in Scenario A), $\ell^p_{\rm avg}= \langle\ell\rangle$, the average path length~\cite{AlbertBarabasi02}. On the other hand, if all links are ME (i.e., $M=L$) in Scenario B, $\ell^p_{\rm avg}=0$, as the entire network can achieve $100\%$ fidelity. Since, in scenarios A and B, we know $F^{\rm max}_{\rm avg}$ as a polynomial in $p$ with path lengths appearing in the exponents, the average of effective path lengths in a network can be related to $F^{\rm max}_{\rm avg}$ in a simple manner:
\begin{align}
    \ell^{p}_{\rm avg} = 2\left(\partial F^{\rm max}_{\rm avg}/\partial p\right)_{p\to1}.\label{eq:lpavg}
\end{align}
Since $F^{\rm max}_{\rm avg}=1$ for $p=1$, we can use this to estimate $F^{\rm max}_{\rm avg}$ for $p$ close to $1$: $F^{\rm max}_{\rm avg}(1-\Delta p)\approx 1-\ell^{p}_{\rm avg}\Delta p/2$.  It is not difficult to generalise Eq.~\eqref{eq:lpavg} to the fully general scenario, Scenario C (where we have $\mathbf{p}=\{p_1,p_2,\ldots,p_L\}$ instead of a single $p$):
\begin{align}
    \ell^p_{\rm avg} = \sum_{i\neq j} \ell^{p_i}_{\rm avg} =&\ \left.2\sum_{i\neq j}\frac{\partial F^{\rm max}_{\rm avg}}{\partial p_i}\right|_{\mathbf{p}\to \mathbf{1}},\\
    F^{\rm max}_{\rm avg}(\mathbf{1}-\Delta \mathbf{p}) \approx&\ 1 - \sum_{i\neq j} \ell^{p_i}_{\rm avg}\Delta {p_i}/2,
\end{align}
where the above sums exclude any index $j$ if $p_j=1$. 
\medskip

\begin{figure}[!t] 
\centering
\includegraphics[width=0.975\columnwidth]{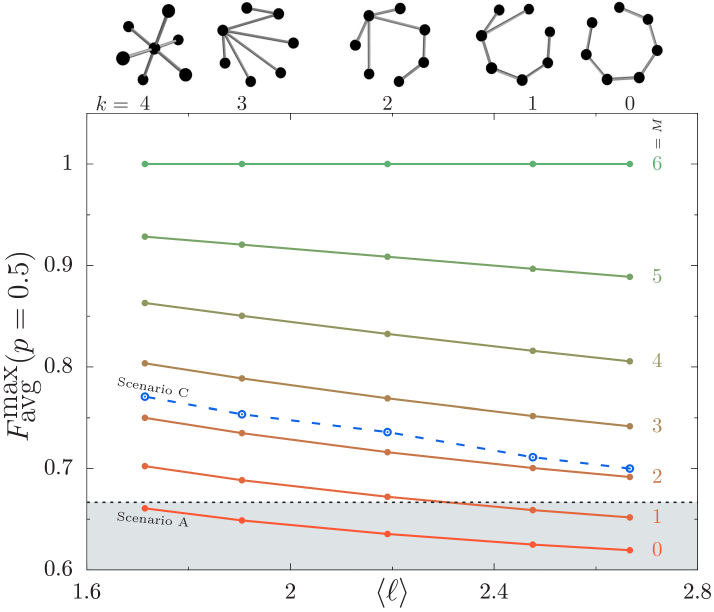} 
\caption{\label{fig:fig2}The role of topology: The average teleportation fidelity, $F^{\rm max}_{\rm avg}$, of seven-node networks constructed sequentially (as illustrated in Fig. \ref{fig:fig1}) with Werner states; $F^{\rm max}_{\rm avg}<2/3$ in the shaded region. The bottom-most red line is for Scenario A ($M=0$), and the other solid lines are for different $M$ values in Scenario B. For illustration, we also show a blue dashed line from Scenario C where the $p_i$'s are randomly drawn from the uniform distribution. }
\end{figure}

\begin{figure}[!t]
\centering
{\includegraphics[scale=0.45]{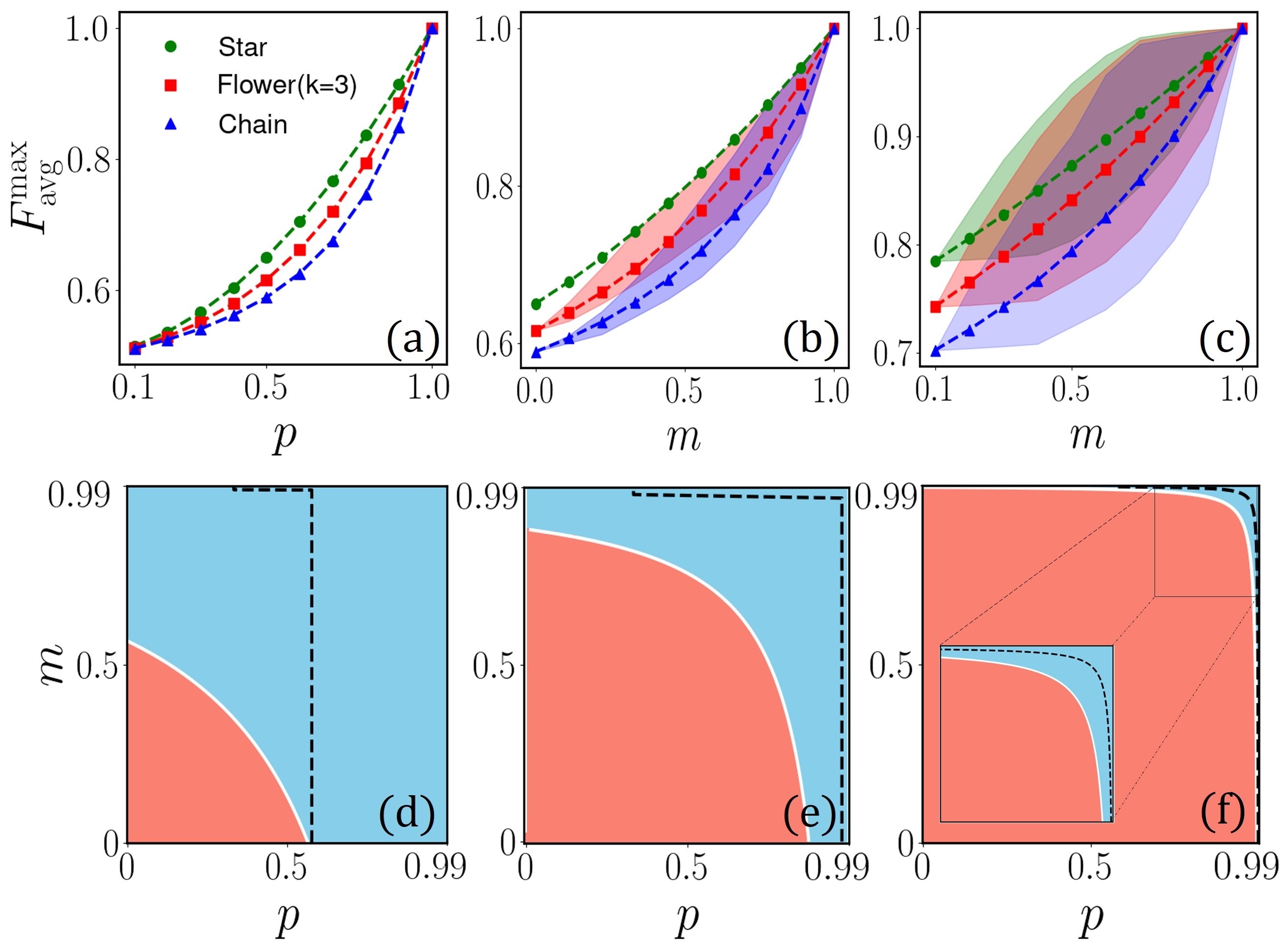}
\caption{\label{fig:Figure_phasespce_mvs_p}(Top panel $N=10$) The dependence of $F^{\rm max}_{\rm avg}$ on $p$ and $m=M/L$ for the chain, the third intermediate flower, and the star in (a) Scenario A, (b) Scenario B (for $p=0.5$), and (c) Scenario C. 
The shaded regions show the effect of permuting the links. (Bottom panel $N=100$) The regime of average quantum advantage (cyan): (d) for the star, (e) the $48^{\rm th}$ intermediate flower (which has $50$ petals), and (f) the chain. To the right of the dashed black lines, every path has $F^{\rm max}>2/3$.}}
\includegraphics[scale=0.65]{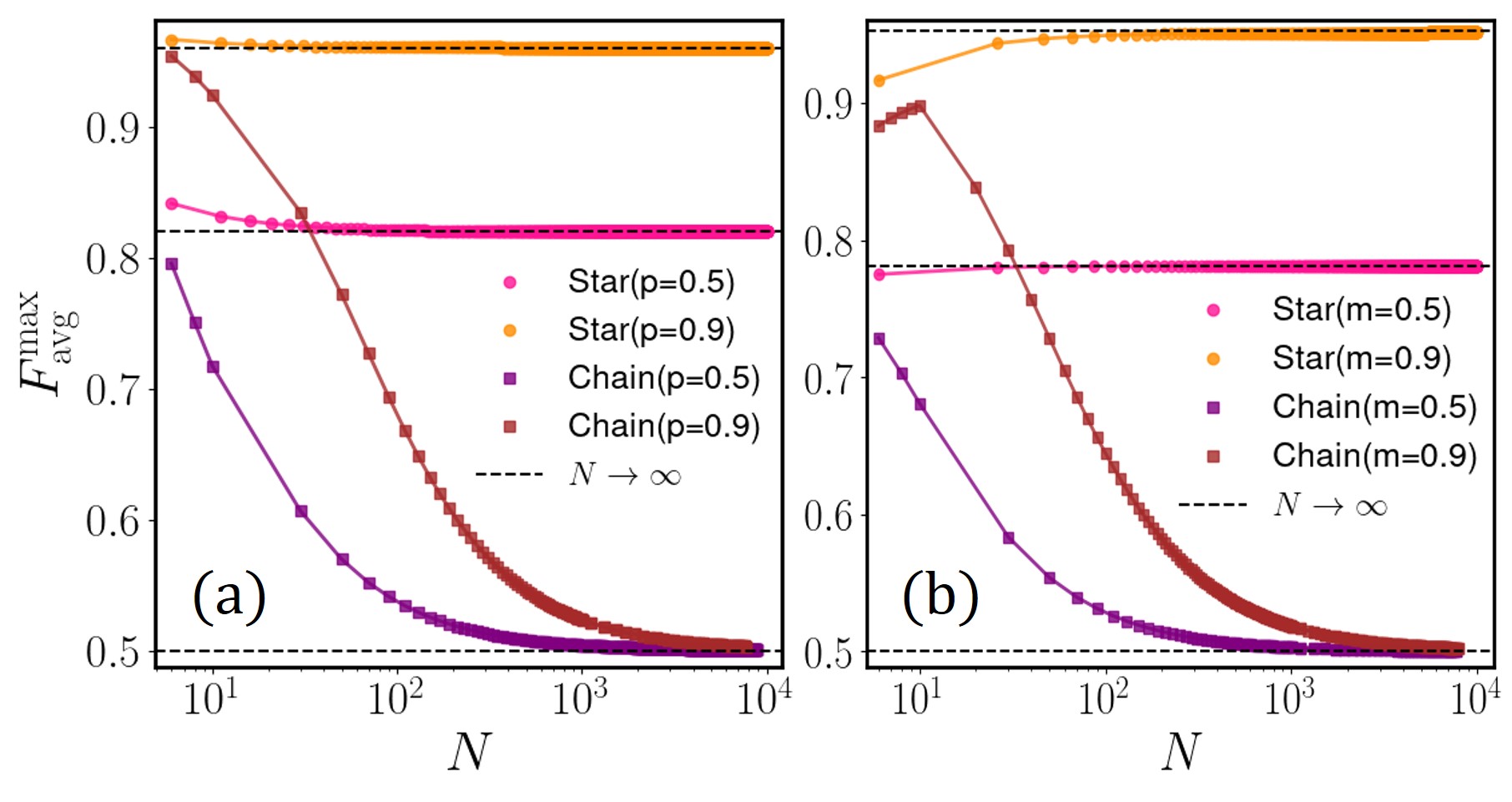}
\caption{The role of network size ($N$) in $F^{\rm max}_{\rm avg}$: We consider four cases in Scenario B: $p=\{0.5,0.9\}$, $m=M/L=0.6$ and $m=\{0.5,0.9\}$, $p=0.5$. For large $N$, as expected, $F^{\rm max}_{\rm avg}$ for the chain always approaches $0.5$ as long as $p,m<1$. The dashed lines show the analytical results and the points show the results of direct numerical estimations.}
\label{fig:Figure_N}
\end{figure}

\noindent \textit{Numerical results:}
For illustration, we show the dependence of $F^{\rm max}_{\rm avg}$ on the average path length, $\langle\ell\rangle$, for the five possible graphs of $7$ nodes and $p=1/2$ in Fig.~\ref{fig:fig2}. As expected, $F^{\rm max}_{\rm avg}$ decreases as we go from the star to the chain. However, on average, $p=1/2$ is insufficient for any graph to achieve quantum advantage as $F^{\rm max}_{\rm avg}< 2/3$ for all topologies of our interest. (This is in contrast to a single link, which can show quantum advantage if $p> 1/3$.) The situation improves with the introduction of ME links. For $2\leq M\leq 6$, all graphs can achieve quantum advantage for the same value of $p$, while for $M=1$, only the second or higher IFs show $F^{\rm max}_{\rm avg}>2/3$.
In Fig.~\ref{fig:Figure_phasespce_mvs_p}, we illustrate the dependence on $p$ and $m=M/L$ for $N=10$ (top row) and $N=100$ (bottom row) chain, star and IFs. 
{Plot (a) shows the variations of $F^{\rm max}_{\rm avg}$ with $p$ in Scenario A. The dots represent simulated results, and the lines are from theory. For all values of $p$, the star performs the best. Plot (b) shows the variations of $F^{\rm max}_{\rm avg}$ with $m~(=M/L)$ for $p=0.5$ in Scenario B (the shaded regions are obtained by shuffling the links). Plot (c) is for Scenario C: for $m$ values, the star outperforms others. In (d), (e), and (f), we consider $N=100$ and mark the regions in the $p-m$ plane with $F^{\rm max}_{\rm avg}>2/3$ for the star, the $48$th IF and the chain, respectively. These plots [(d)-(f)] are obtained by averaging over possible arrangements of the ME links. We show the large-$N$ limits in Fig.~\ref{fig:Figure_N} for two benchmark choices of $m$ and $p$ in star and chain: $\{0.5,0.9\}$. With the increase in the number of large paths, $F^{\rm max}_{\rm avg}\to1/2$ in chains for any value of $m<1$, as expected.}
\medskip

\noindent \textit{Summary and conclusions:} In this letter, we studied large quantum repeater networks with Werner states-based models. These simple models let us analyse the achievable teleportation fidelities ($\sim$ abilities to transfer quantum information) of complex repeater-based networks with a few parameters (e.g., weights of Werner states, $\{p_i\}$, the network size, $N$, etc.). We considered three scenarios where not all links in a network were maximally entangled, as one would expect in a practical setup: (A) all $p_i=p$ with $0<p<1$; (B) a fraction of the links are maximally entangled while all others have $p_i=p$; and (C) the $p_i$'s are randomly sampled from the uniform distribution. In these scenarios, we characterised networks of various loopless topologies (chain to star) in terms of their average maximum fidelities (average of the maximum fidelities between all pairs of nodes), $F^{\rm max}_{\rm avg}$. It is a measure of the resourcefulness of a network as a whole (i.e., a global/typical measure), as it is independent of the choice of source and target nodes. The fidelity of a network is $100\%$ when all its links are ME states. However, because of factors like noise and ageing, all states may not be ME in practice. In such situations, $F^{\rm max}_{\rm avg}$ is useful to compare networks across topologies. 

We quantitatively showed how, for a fixed network size, $F^{\rm max}_{\rm avg}$ increased with the degree of the network: minimum for the chain and maximum for the star {(among loopless networks)}. Besides these two extreme topologies, we also obtained analytic expressions of $F^{\rm max}_{\rm avg}$ for the IFs (which are representative trees of the same size) in Scenario A and B.\footnote{The IFs are representative since $F^{\rm max}_{\rm avg}$ for all intermediate trees will lie between that for the chain and the star.} We estimated the parameter values for which a network as a whole is expected to show quantum advantages, i.e., show $F^{\rm max}_{\rm avg}>2/3$ (this is impossible if no path has any entangled state). For large networks, the $p$ value at this threshold depends on the network topology. We see that no chain can show quantum advantages in the large $N$-limit as, $F^{\rm max}_{\rm avg}\to 1/2$ for $m<1$. However, a star can, as long as $p>1/\sqrt{3}$. 
We also found an interesting relationship between the derivative of $F^{\rm max}_{\rm avg}$ and $\ell^p_{\rm avg}$, the average effective path lengths, which is essentially the average of the resistance distance of the network. It allows for estimating $F^{\rm max}_{\rm avg}$ for $p_i$ close to $1$ without performing any measurements by simply drawing an equivalent resistance network and calculating the resistance distance.  

{So far, we have restricted our discussions to loopless networks. However, the $F^{\rm max}_{\rm avg}(\rho_{\rm wer})$ defined in Eq.~\eqref{eq:fmaxavgdef} applies to networks with loops as well. We demonstrate this with the explicit examples of the simple ring and the complete graph in Appendix~\hyperref[sec:loops]{B}. Topologies with loops must be analysed carefully, as multiply-connected nodes generally lead to subtleties. We show the analytical expressions of $F^{\rm max}_{\rm avg}(\rho_{\rm wer})$ in Scenario A for both the complete graph and ring to understand the fidelity bounds. We plan to explore the non-trivial effects a complicated loop can show in detail in a future publication. Apart from teleportation, there could be similar measures for other tasks such as secure communication, blind quantum computation, secret sharing, distributed sensing, etc. Investigating how to define similar global metrics for quantifying resourcefulness and rank networks for these tasks will be interesting. Some multi-party extensions~\cite{meignant2019distributing} of the existing work might become necessary in some cases.}
\medskip

\begin{acknowledgments}
\noindent\textit{Acknowledgements:} We thank Damian Markham, Siddhartha Das, and Harjinder Singh for their helpful comments on the draft.\\ 
\end{acknowledgments}
\input{avgmaxfidelity_main.bbl}
\relax

\onecolumngrid
\section*{End Matter}

\section{Appendix A. Intermediate flowers in Scenario B}\label{sec:appscenarioB}
\renewcommand{\theequation}{A.\arabic{equation}}
\setcounter{equation}{0}
\noindent
If we assume $m_s$ of the $\ell_s=k+2$ links connecting the petals of the $k^{\rm th}$ IF are ME, we get 
\begin{align}
 \left.\hspace{-0.1cm}F^{\rm max}_{ \rm avg}(N,M,p)\right|_{{\rm flower}_k}=&\ \frac{1}{^{N}\!C_2\, ^{L}\!C_M}\sum_{m_s}\Bigg[ 
 {^{\ell_s}\!C_{m_s}}{^{\ell_c}\!C_{m_c}}\Bigg\{\Bigg({^{m_s+1}\!C_2}\mathcal F_0 + (m_s+1)(\ell_s-m_s)\mathcal F_1 + {^{\ell_s-m_s}\!C_2}\mathcal F_2\Bigg)+\Bigg(\frac{m_c(\ell_c+1)}{(\ell_c+2-m_c)}\mathcal F_0\nonumber\\
 &\ + \Big(\prod_{i=1}^2\frac{(\ell_c+i)}{(\ell_c+i-m_c)}\Big) \sum\limits_{\ell = 1}^{\ell_c-m_c}(\ell_c+1-m_c-\ell)\mathcal F_\ell\Bigg)\Bigg\}
 +
 \Bigg\{
 \sum_{i=1}^{\ell_c}\sum_{\ell=i-m_c}^{\ell_c-m_c}\Bigg((\ell_s-m_s-1){^{\ell_s-1}\!C_{m_s}}\mathcal F_{(\ell+2)} \nonumber\\
 &\ +\Big((\ell_s-m_s){^{\ell_s-1}\!C_{m_s-1}}+(m_s+1){^{\ell_s-1}\!C_{m_s}}\Big)\mathcal F_{(\ell+1)}+ m_s{^{\ell_s-1}\!C_{m_s-1}}\mathcal F_{\ell}
 \Bigg){^{i}\!C_{\ell}}\,{^{\ell_c-i}\!C_{\ell_c-m_c-\ell}}\Bigg\}
 \Bigg],
  \label{eq:kthif_m}
\end{align}
where $\ell_c = L-\ell_s$ and $m_c=M-m_s$, and the $m_s$ sum runs over all possibilities such that $0\leq m_s\leq l_s, M$ and $0\leq m_c\leq l_c, M$. In the above equation, we have grouped the terms so that the star, chain, and overlap contributions can be identified easily.
\section{Appendix B. Ring and the complete graph: $F_{\rm avg}^{\rm max}$ for graphs with loops}\label{sec:loops}
\renewcommand{\theequation}{B.\arabic{equation}}
\setcounter{equation}{0}
\noindent
{In the main text, we have focused only on loopless networks for simplicity. However, it is easy to see that the definition, $F^{\rm max}_{\rm avg}= \langle F^{\rm max}_{{\rm ST},\mathcal P_{\rm max}}\rangle_{{\rm ST}}$, in Eq.~\eqref{eq:fmaxavgdef} also applies to topologies with loops. For example, we can consider two extreme cases---a simple ring (i.e., a chain with two ends connected) and a complete graph (where there is a direct link between each pair of nodes). In the ring, every two nodes are connected by (at least) two paths. Hence, for every pair of S and T, we identify the path giving the best fidelity ($\mathcal P_{\rm max}$) to calculate the average of maximum achievable fidelity. We can also similarly estimate it for the complete graph, where more paths exist between a pair of nodes.} 

\begin{table}[!t]
\caption{{The average of the maximum achievable fidelity, $F^{\rm max}_{\rm avg}(\rho_{\rm wer})$, for four basic topologies with $N=4$. The Scenario C numbers are obtained by averaging over a large number of random samples drawn from the uniform distribution, $\mathcal U_{[0,1]}$.\label{tab:1}}}
\centering
{
\renewcommand\baselinestretch{1.25}\selectfont
\begin{tabular*}{0.5\columnwidth}{l@{\extracolsep{\fill}} cc}
\hline
    \multirow{3}{*}{Topology}&\multicolumn{2}{c}{$F^{\rm max}_{\rm avg}(\rho_{\rm wer})$}\\\cline{2-3}&Scenario A &Scenario C\\
    &($p=1/2$)&$(p_i\in\mathcal U_{[0,1]})$\\
    \hline \hline
    Chain&$65/96$&$0.6771$\\
    Star&$66/96$&$0.6875$\\
    Ring&$66/96$&$0.7300$\\
    Complete&$72/96$&$0.8000$\\
    \hline
    \end{tabular*}
    }
\end{table}

{For an illustration, we show $F^{\rm max}_{\rm avg}$ for the basic topologies in Table~\ref{tab:1}. It is easy to understand the results in Scenario A as, in this case, the shortest paths achieve the maximum fidelity since $p^2<p$ (unless $p=1$). For example, every S and T combination contributes $(1+p)/2$ to $F^{\rm max}_{\rm tot}$ in the complete graph and together gives 
\begin{equation}
\left.F^{\rm max}_{\rm avg}(p)\right|_{\rm complete}=(1+p)/2. 
\end{equation}
Similarly, there are always two paths between any two nodes in a ring. Selecting the shorter paths, we get,
\begin{equation}
\left.F^{\rm max}_{ \rm avg}(N,p)\right|_{\rm ring}=\frac{1}{ \lfloor{N/2} \rfloor }\left[\sum_{\ell=1}^{ \lfloor{N/2} \rfloor}\mathcal F_\ell\right].
  \label{eq:p_ring}
\end{equation}
where $\lfloor x\rfloor$ is the floor function, giving the integer part of the argument. Note that when $N$ is even, a pair of opposite nodes is connected by two equal-length paths, both contributing equally to $F^{\rm max}_{ \rm tot}$. Eq.~\eqref{eq:p_ring} takes this degeneracy into account.}

{For loopless topologies, where the path between any S and T pair is unique, both Scenario A and C lead to the same numbers because the average of the uniform distribution is $0.5$. However, we see something interesting when there is more than one path. For the ring and the complete graph, we first calculate the maximum fidelity for every S and T pair and then calculate the average (over different S and T combinations and samples), but these two operations do not commute. As a result, the ring and complete graph have larger $F^{\rm max}_{\rm avg}$ than those in Scenario A because a longer path can achieve higher fidelity than a shorter path if the $p_i$'s are chosen randomly. For an intuitive understanding, explicitly considering the example of $N=3$ loop (where the complete graph is the same as the ring) is perhaps the best. In this case, there are three links with weight factors $p_1$, $p_2$, and $p_3\in \mathcal U_{[0,1]}$. We get,
\begin{align}
    \left.F^{\rm max}_{\rm avg}(3)\right|_{\rm complete/ring}
    &= \int_0^1\int_0^1\int_0^1 dp_1dp_2dp_3\ {\rm Max}\left(\frac{1+p_1}{2},\frac{1+p_2p_3}{2}\right)\nonumber\\
    &=\int_0^1 dp_1\int_{p_1}^1 dp_2\int_{\frac{p_1}{p_2}}^1 dp_3 \frac{1+p_2p_3}{2}
    \quad +\int_0^1 dp_1 \frac{1+p_1}{2}\left(1-\int_{p_1}^1 dp_2\int_{\frac{p_1}{p_2}}^1 dp_3 \right),
\end{align}
where $p_2p_3>p_1$ in the first term of the last line. The above integral evaluates to $7/9$, which is larger than $3/4$, the answer we would have got if the maximisation were performed after the integrations. We do not see this effect for loopless topologies because there, the path between any S and T pair is unique. The above illustration establishes two points we made in the letter. First, even with loops, $F^{\rm max}_{\rm avg}$ behaves as expected---as it is always larger for the complete graph than the ring. Second, loops show non-trivial effects and hence require a separate dedicated analysis.} 

\section{Appendix C. The effect of decoherence}\label{sec:decoh}
\renewcommand{\theequation}{C.\arabic{equation}}
\noindent
{We want to see how the intra-node distance can affect $F^{\rm max}_{\rm avg}(\rho_{\rm wer})$. As a concrete example, let us consider the case of photonic qubits. In this case, we can simply model the probability $p$ associated with the Werner (depolarised) state as the success probability of a qubit to remain entangled after transfer~\cite{da2024requirements}. This is possible even if we start with maximally entangled states. As the qubit passes through a fibre, its coherence declines as a function of the distance $d$ it covers inside. As a result, we can parametrise the resultant shared states with $p$.}

\begin{figure}[b!]
\centering
\includegraphics[scale=0.45]{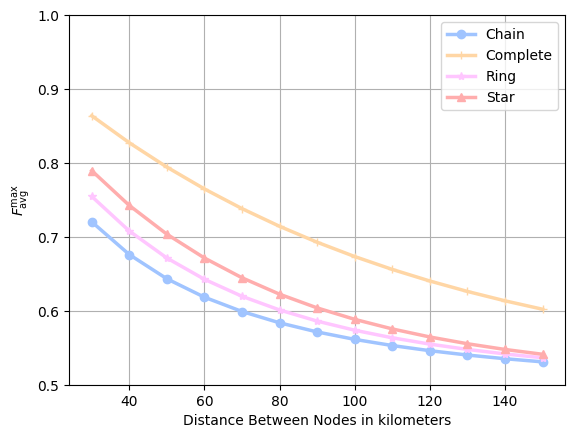}
\caption{\label{fig:decohe}The average of maximum fidelity, $F^{\rm max}_{\rm avg}(\rho_{\rm wer})$, for four basic topologies with $N=8$ is plotted against the distance between nodes.}  
\end{figure}

{Let $t=d/c$ be the time required to have one attempt to transfer a qubit, where $d$ is the fibre distance between two nodes and $c$ is the speed of light in the fibre. (It could be the communication time of sending a photon to the heralding station.) The probability of having one successful attempt can be modelled as~\cite{da2024requirements},
\begin{equation}
    p_{s} = p_{\rm det}\times 10^{-\alpha d / 10},
\end{equation}
where $\alpha$ is the fibre attenuation coefficient and $p_{\rm det}$ is the probability that the photon is detected by a detector. The photon is detected means $p_{\rm det}=1$. Then, $p_{s} = 10^{-\alpha d / 10}$ gives the success probability of establishing an entangled link. We model the success probability $p_s$ by the Werner state probability $p$ (representing the component of the Bell state), i.e., $p_s\equiv p$. In Fig.~\ref{fig:decohe}, we plot $F^{\rm max}_{\rm avg}(\rho_{\rm wer})$ against the distance between each node for all the four topologies to find the loss is maximum for chain and minimum for a complete graph. As expected, with the increasing distance between the nodes $F^{\rm max}_{\rm avg}(\rho_{\rm wer})$ goes down. Here $\alpha=0.46$, $p_{\rm det}=1$, and we have taken the number of nodes to be $8$, $d$ value ranges from $30$ km to $150$ km, incremented by $10$ km for each run.}

{This illustrates the successful modelling of one real noise parameter by the parameter $p$ associated with the Werner state.}

\end{document}

%% file: avgmaxfidelity_main.bbl
%